\documentclass{amsproc}

\theoremstyle{definition}

\theoremstyle{remark}

\numberwithin{equation}{section}

\begin{document}

\title{New types of solvability in {\it PT}  symmetric quantum theory}

\author{Miloslav Znojil}
\address{\'{U}stav jadern\'e fyziky AV \v{C}R, 250 68 \v{R}e\v{z}, Czech
Republic }
\email{znojil@ujf.cas.cz}
\thanks{The author was supported in part by AS CR Grant \#A1048004.}


\date{November 15, 2002.}



\begin{abstract}

The characteristic anti-linear (parity/time reversal, PT) symmetry of
non-Hermitian Hamiltonians with real energies is presented as a
source of two new forms of solvability of Schr\"{o}dinger's
bound-state problems. In detail we describe (1) their very specific
semi-exact solvability (SES) and (2) their innovated variational
tractability. SES technicalities are discussed via charged oscillator
example. In a broader context, speculations are added concerning
possible relationship between PT symmetry, solvability and
superintegrability.

\end{abstract}
\maketitle

\section*{${\it PT}$ symmetric quantum mechanics: A brief introduction}

\subsection*{Prehistory: Isospectral operators}

Origins of the popular Bender's and Boettcher's ${\it PT}$ symmetric
quantum mechanics \cite{BB,BBM} lie in perturbation theory
\cite{Caliceti,Calicetib}. For an elementary illustration one may
recollect the pioneering paper by Buslaev and Grecchi \cite{BG} who
proved the isospectrality of the Hermitian, spherically symmetric
$D-$dimensional perturbed harmonic oscillator
 \[
 H^{(PHO)}(g) = \frac{1}{2}\left (
 -\triangle + \sum_{j=1}^{D}x_j^2
 \right )+
 g^2\left (
  \sum_{j=1}^{D}x_j^2
 \right )^2\
 \]
(in its $m-$th partial-wave projection at any $m = 0, 1,\ldots$) with
its {\em non-Hermitian} ``unstable" anharmonic partner(s) in one
dimension,
 \[
   H^{(UAO)}(g)=
 -\frac{d^2}{dz^2} + z^2 (i g z-1)^2-(D +2m - 2)\,(i g z-1/2)\,
  , \ \ \ \ \ \ g > 0.
 \]
This means that the physics (and, in particular, the reality of
energies) remains unchanged while the mathematics itself is
significantly simplified when one weakens the Hermiticity of
$H^{(PHO)}(g)=\left [H^{(PHO)}(g)\right ]^\dagger$ to the mere ${\it
PT}$ symmetry of its spectrally equivalent partners \cite{DDT}. The
latter operators commute with the product ${\it PT}$ of parity and
time reversal. In the modern language of review \cite{Mostafazadeh},
one should rather speak about the pseudo-Hermiticity defined by the
relation
 \[
 H^{(UAO)}(g)= P\,\left [ H^{(UAO)}(g)\right ]^\dagger \,P.
 \]

\subsection*{Recent progress: Exactly solvable examples }

Paper \cite{BG} extends the proof to the isospectrality between
$H^{(PHO)}(g)$ and $H^{(UAO)}(g)$ to the {imaginary} couplings
$g=ih$. The spectrum of the resulting self-adjoint double-well
$H^{(UAO)}(ih)$ in one dimension (famous for its perceivably hindered
perturbative tractability) may be then deduced from the ``unstable"
modification $H^{(PHO)}(ih)$ of the central quartic oscillator in $D$
dimensions. The latter, non-Hermitian ${\it PT}$ symmetric model is
sufficiently simple in its ordinary differential ``radial"
representation of ref. \cite{BG},
 \[
  H_{\varepsilon}^{(URO)}(h)=
 -\,\frac{d^2}{dr^2}+ \frac{\ell(\ell+1)}{r^2}+ r^2-h^2\,r^4
 ,
 \]
 \[
 \ \ \ \ \
 \ \ \ \ \
 \ \ \ \ \
 r = r(x) = x -i\,\varepsilon,
  \ \ \ \ \ \ \  x \in (-\infty, \infty),
   \ \ \ \ \ \ h > 0,\ \ \ \varepsilon > 0.
 \]
In the context of the present proceedings this operator admits an
immediate re-interpretation as a specific perturbed and regularized
form of the most common superintegrable oscillator $H_0^{(URO)}(0)$
of Smorodinnsky and Winternitz (SW, \cite{SW,SWE}) in any number of
dimensions and after its separation in cartesian coordinates.

The latter observation is inspiring since it is known that the
concepts of (maximal) superintegrability and (exact) solvability are
closely related at $\varepsilon = 0$ \cite{TTW}. In such a setting
the weakening $\varepsilon>0$ of the Hermiticity extends the class of
the solvable models \cite{SI,SIb} as well as their interpretation in
terms of Lie algebras \cite{Levai} and evokes a number of new open
questions. We have addressed the first few of them in ref.
\cite{ptho} where we derived the exact spectrum and wave functions
for the complexified SW oscillator $ H_{\varepsilon}^{(URO)}(0)$. Our
next move in this direction was devoted to the closely related
complexified (though still separable) Calogero model of three
particles \cite{Tatera} where an overlap of the integrability with
${\it PT}$ symmetry acted as a generator of the new solvable
self-adjoint Hamiltonians \cite{Taterb}. This experience makes the
further study of ${\it PT}$ symmetric models of (super)integrable
type very promising. In the present paper, we are going to describe
some new results achieved in this direction.

\subsection*{Running project: Partially solvable models }

In place of the (not so easily feasible) analysis of the traditional
quartic perturbations, we shall pay attention to the slightly simpler
charged oscillators characterized, first of all, by their
incompletely or quasi-exactly solvable\footnote{term coined by A.
Turbiner, meaning an elementary solvability for finite multiplets of
states} (QES) status. From the historical perspective, the acceptance
of this concept was comparatively dramatic. In a prelude, various
potentials have been shown partially solvable in terms of elementary
functions.  Thirty years ago, this period has been initiated by a
two-page remark by Andr\'{e} Hautot \cite{Hautot} who noticed that
the charged oscillator possesses arbitrary finite multiplets of exact
Sturmian solutions in two ($D=2$) and three ($D=3$) dimensions. The
Hautot's choice of the elementary wave-function ansatz
 \[
\varphi(r) = e^{-r^2/2-g\,r}\ \sum_{n=0}^{N}\ h_n \,r^{n+\kappa}\
\label{ransatz}
 \]
(with arbitrary $N$) preceded the discoveries of the QES sextic
oscillator \cite{Singh}, of non-polynomial QES anharmonicities
\cite{Flessas} etc. During the ``golden age" of the development of
the subject, the existence of the common Lie-algebraic background of
all the QES systems has been emphasized \cite{Turbiner}. A summary of
the ``state of the art" up to the early nineties has been offered by
Alex Ushveridze in his monograph \cite{Ushveridze} where the quartic
polynomial oscillator is mentioned as a typical system {\em without}
quasi-exact solvability.

A change of the approach to QES models has been initiated by Bender
and Boettcher \cite{BBjpa} who revealed and described the QES
solutions for the ``unstable" quartic polynomial oscillators. Their
construction has been extended to all the partial waves in ref.
\cite{quartic}. In paper \cite{I} we made the next move and turned
attention to ${\it PT}$ symmetrization of the Hautot's charged
oscillator (cf. Appendix A). In the present continuation of this
effort, we shall complete the picture by its extension, i.a., to all
the partial waves (cf. Appendix B). In section 1 we describe our main
result, viz., the quasi-even solutions which remained unnoticed in
\cite{I}. Unexpectedly, these ``facilitated QES" states are available
in {\em infinite} multiplets\footnote{such a feature is much more
characteristic for the completely solvable models}, all the elements
of which are, basically, non-numerical. In a way emphasized in
section 2, these sets may easily serve as certain non-standard bases
in some ``sufficiently large" subspaces of Hilbert space. For this
reason, we suggest to call them semi-exactly solvable (SES, cf.
Table~1).

\begin{table}[ht]
\caption{Tentative classification of solvability.}\label{table2}
\renewcommand\arraystretch{1.5}
\noindent\[
 \begin{array}{||c||c|c|c|c|c|c|c||} \hline
 \hline
 {\rm class}
 &\multicolumn{2}{c|}{\rm
 quasi-exact}
 &\multicolumn{2}{c|}{\rm semi-exact }
 &\multicolumn{2}{c|}{\rm exact }
 \\
 \hline
  \multicolumn{1}{||c||}{\rm  solutions\ available }
 &
\multicolumn{2}{|c|}{{\rm  finite\ set}} & \multicolumn{2}{|c|}{{\rm
 infinitely\ many}}
   &
\multicolumn{2}{c|}{{\rm  all} }
 \\
 \multicolumn{1}{||c||}{\rm range\ of\ couplings
  }
  & \multicolumn{2}{|c|}{\rm
 restricted}
  &
 \multicolumn{2}{|c|}{\rm restricted}
  &\multicolumn{2}{|c|}{\rm any}
 \\
 \multicolumn{1}{||c||}{\rm illustrative\ example  }
 & \multicolumn{2}{|c|}
 {H^{(BBO)}(a,b) }
  &
 \multicolumn{2}{|c|}{\rm  see\ below}
  &\multicolumn{2}{|c|}{H_{\varepsilon}^{(URO)}(0) }
 \\
   \hline
\hline
\end{array}
\]
\end{table}

\specialsection{Quasi-exact solvability on complex contours}

\subsection{Three-term recurrences}

Let us consider the Schr\"{o}dinger equation
 \begin{equation}
 \left[-\,\frac{d^2}{dx^2}+
 \frac{\ell(\ell+1)}{r^2(x)}+ i\,\frac{F}{r(x)} +
 2\,i\,b\,r(x) +  r^2(x)\right]\,
 \psi_n(x) = E_n\, \psi_n(x)
 \label{SE}
 \end{equation}
 \[
 \ell=\ell({\it  L})=({\it  L} - 1)/2, \ \ \ \ \ \ \ \
 {\it  L} = D-2, D, D+2,\ldots\ .
 \]
It contrast to the Hautot's QES problem of Appendix A it works with
the purely imaginary charge and with a constant complex shift of
coordinates $r = r(x) = x -i\,\varepsilon$ (cf. \cite{Ahmed}). It
also generalizes the $\ell=0$ problem of ref. \cite{I} so that we
must employ the more powerful ansatz
 \begin{equation}
 \psi_{n}(r) = e^{-r^2/2-i\,b\,r}\
\sum_{n=0}^{N} \ (i\,r)^{n-\ell}\ p_n
  \label{ansatz}.
 \end{equation}
After its insertion, the differential form of our ${\it PT}$
symmetric bound state problem (\ref{SE}) is replaced by the
finite-dimensional matrix equation,
\begin{equation}
 \left( \begin{array}{ccccc} B_{0} & C_0&  & &  \\
A_1&B_{1} & C_1&    & \\
 &\ddots&\ddots&\ddots&\\
&&A_{N-1}&B_{N-1}&C_{N-1}\\
 &&&A_{N}&B_{N}\\
 &&&&A_{N+1}
\end{array} \right) \left ( \begin{array}{c} p_0\\
 p_2\\
 \vdots \\
 p_N \end{array} \right)
= 0\ , \ \ \ \ \ \ N < \infty
 \label{recu}
 \end{equation}
with the {\em real} matrix elements,
 \[
\begin{array}{c}
 A_n =b^2+ 2n  -{\it L}-E, \ \ \ \ \ \ \
 B_n =  -(2n+1-{\it L})b-F,\\
 C_n =  (n+1)\,(n+1-{\it L}),\ \ \ \ \ \ \
  {\it L} = 2\ell+1, \ \  \
  \ \ \ \ n = 0, 1, \ldots \ .
  \end{array}
   \]
The last line of (\ref{recu}) represents a separate condition
$A_{N+1}=0$ which gives our energies in closed form,
 \begin{equation}
 E=
 E_N = 2N+2-{\it L} +b^2,\ \ \ \ \ \ \ \ \ \
 N = 0, 1, \ldots \, .
 \label{Hautots}
 \end{equation}
The key consequence of this easy but important simplification lies in
the emergence of a zero $C_{L-1}=0$ in the upper diagonal of the
square matrix of the simplified system (\ref{recu}). In Appendix~B
the presence of this zero enables us to construct the quasi-odd
solutions at all~$\ell$.

\subsection{The new, quasi-even solutions}

All the coefficients in the polynomial wave functions (\ref{ansatz})
may be defined by closed formula~(\ref{exrecu}) displayed in
Appendix~B. Let us now turn our attention to the quasi-even states.
The superscript $^{(+)}$ will be introduced to mark their even
quasi-parity which may be characterized, for our present purposes, by
the simple negation of the odd quasi-parity condition~(\ref{condi}),
\begin{equation}
|p_0^{(+)}| + |p_1^{(+)}| +\ldots + | p_{{\it L}-1}^{(+)}| > 0.
\label{condibe}
 \end{equation}
For quasi-even states, the vanishing matrix element $ C_{{\it
L}-1}^{(+)}=0$ plays a different role. It separates again the two
subsets of equations but the upper one ceases to be trivial. This
means that the related subdeterminant must vanish,
 \begin{equation}
 \label{recupe}
  {\it S}^{(+)} =\det
 \left( \begin{array}{ccccc} B_{0}{{}}
  & C_{0}{{}}
 &  & &  \\
 A_{1}{{}}&B_{1}{{}} & C_{1}{{}}&
 &
 \\
 &\ddots&\ddots&\ddots&\\
 &&A_{{\it L}-2}{{}}
 &B_{{\it L}-2}{{}}&C_{{\it L}-2}{{}}\\
  &&&A_{{\it L}-1}{{}}&B_{{\it L}-1}{{}}
 \end{array}
  \right)\ = 0.
 \end{equation}
Up to the exceptional, degenerate cases where eqs. (\ref{recupe}) and
(\ref{qesrec}) hold at the same time, we may drop the non-vanishing
factor $ {\it S}^{(-)}\neq 0$ and determine all the charges
$F=F^{(+)}_{N,k}$ as roots of polynomial (\ref{recupe}).

The $^{(+)}-$superscripted wave function coefficients are determined
by eq. (\ref{exrecu}), this time in the full range of indices $j = N,
N-1, \ldots, 1 $. As long as the dimension of these determinants
grows with the difference $N-j$, we shall recommend a return to the
recurrences
 \[
 \left( \begin{array}{cccccc}
-2N&\beta_1-F & 4-2{\it L}&    & &\\ &2-2N&\beta_2-F & 9-3{\it L}& &
\\
 &&\ddots&\ddots&\ddots&\\
&&&-4&\beta_{N-1}-F&N^2-N\,{\it L}
\\
& &&&-2&\beta_N-F
 \end{array} \right)
 \left ( \begin{array}{c}
p_0^{(+)}\\ p_1^{(+)}\\ \vdots
\\
 p_{N-1}^{(+)}
  \\
 p_N^{(+)} \end{array} \right)
= 0\
 \]
to be read from below upwards\footnote{we abbreviated $\beta_n \equiv
-(2n+1-{\it L})\,b$ for the sake of brevity}. The process must be
initiated at $p_N^{(+)} \neq 0$. The vanishing $C_{{\it L}-1}=0$ does
not play any role in it.

\section*{Roots of the SES secular determinants  \label{redbenos}}

\subsection{Dimension-independence}

In every partial wave, the size $L$ of the matrix in eq.
(\ref{recupe}) coincides with the degree of the resulting secular
polynomial. This value does not change with the growth of the
dimension $N+1$ of the vector $\vec{p}$. Thus, the degree $N$ of our
polynomial wave functions enters our tridiagonal secular equation as
a mere parameter. Containing extremely elementary matrix elements,
the fully explicit form of this equation reads
 \[
\det \left( \begin{array}{ccccc} ({\it L}-1)b-F & 1-{\it L}&  & &
\\ -2N&({\it L}-3)b-F & \ddots&    & \\ &-2N+2& \ddots & 4-2{\it
L}&\\ &&\ddots &(3-{\it L})b-F&1-{\it L}\\ &&&-2(N+2-{\it L})&(1-{\it
L})b-F
\end{array} \right)
 =0
 \]
and specifies the family of the admissible charges $F =
F^{(+)}_{N,k}$ at any integer $L>0$ and index $k =1, 2, \ldots, {\it
L} $. The first few partial waves are exceptional since their
eigencharges may in principle be defined by closed formulae at ${\it
L}=1$, ${\it L}=2$, ${\it L}=3$ and ${\it L}=4$. At the simplest
choice of ${\it L} = 1$ (which may mean {both} the $s-$wave in three
dimensions {and} an even state at $D=0$), one does not obtain
anything new. Secular equation (\ref{recupe}) provides the single
root $F^{(+)}_{N,1}$ which is equal to zero at all $N$. In the limit
$\varepsilon \to 0$, the quasi-even ${\it L}=1$ solutions converge to
the well known Hermite polynomials. One just re-discovers the exactly
solvable harmonic oscillator basis at even parity. We must
re-emphasize that at any $L\geq 0$ the number of our quasi-even roots
{\em does not change} with the growth of the dimension $N$. This is
of paramount importance since the practical determination of
eigencharges is performed {for all the indices $N$ at once}. Each of
these families numbered by $k$ contains {\em infinitely many}
elements numbered by the first subscript. This might facilitate their
future applications (cf. section 2 below).

\subsection{The first nontrivial generalization of
oscillator basis: ${\it L} = 2$}

Let us move to the index ${\it L}=2$ giving the $p-$wave in two
dimensions or to the $s-$wave in four dimensions. The related
$\ell=1/2$ wave functions may be chosen as compatible with the
even-quasi-parity criterion (\ref{condibe}),
 \[
 p_{0}^{(+)}(N,k) = -\frac{1}{2N}\,
  \left [ {b+F_{N,k}} \right
 ] \,p_{1}^{(+)}(N,k), \ \ \ \ \ \  k = 1, 2\ .
 \]
Secular equation (\ref{recupe}) reads
 \[
\det \left (
\begin{array}{cc}
b-F&-1\\ -2N&-b-F \end{array} \right ) = F^2-b^2-2N=0
 \]
and has two roots
 \begin{equation}
 F^{(+)}_{N,1}=\sqrt{\left( b^{2}+2N\right)}, \ \ \ \ \ \
 F^{(+)}_{N,2}=- \sqrt{\left( b^{2}+2N\right)}
 \label{refex}.
 \end{equation}
[Note the contrast with the necessity of searching for roots of the
polynomial (\ref{qesrec}) of the $N-{\it L}+1$st degree!] We
encounter the non-vanishing charges for the first time. They grow
with the increasing size of the shift $b$ and with the quantum number
$N$ (i.e., with the energy). One finds their set more similar to the
``$F=0$ line" of the harmonic oscillator than to the QES quasi-odd
roots of Appendix B. Indeed, in the latter case the roots
$F^{(-)}_{N,k}$ must be determined by the method which depends
on~$N$. In this sense we may now speak about the most natural and
unique non-Hermitean generalization of the harmonic oscillator
even-parity basis of $L_2(0,\infty)$.

\subsection{Cardano charges at ${\it L} = 3$}

At ${\it L}=3$ the comparatively compact form of our secular equation
\begin{equation}
  \det
\left (
\begin{array}
[c]{ccc}%
2b-F & -2 & 0\\
 -2N & -F & -2\\
 0 & -2N+2 & -2b-F
\end{array}
\right ) = 0
 \label{ouk}
 \end{equation}
enables us to search for the triplet of charges $\{F_{N,1},
F_{N,2},F_{N,3}\}$ via the closed (so called Cardano) formulae. One
of alternative strategies may consist in a transition from the SES
eigencharges $F=F_{N,k}(b)$ to the inverse functions $b =
b_{N',k'}(F)$. Such a trick lowers (by one) the degree of the secular
polynomial at the odd values of ${\it L}$ and, hence, extends the
solvable class up to ${\it L}=5$. A serious shortcoming of such an
approach may be seen in the parallel deformation of the energies
(\ref{Hautots}) which would change with the shift $b$.

In the similar spirit, another simplification of formulae may be
based on the formal elimination of $N$. The consequences may be
illustrated by the secular ${\it L}=3$ determinant (\ref{ouk}) which
represents the mere linear problem for $N$, with the unique solution
 \begin{equation}
 N=-\frac{1}{8F} \left({4Fb^{2}+8b-F^{3}-4F}\right )
 =
 \frac{1}{2}
 \left [
 \left (
 \frac{1}{F} + \frac{F}{2} \right )^2 -
 \left (b + \frac{1}{F} \right )^2\right ]\ .
 \label{ouh}
 \end{equation}
This formula suggests a simultaneous tuning of both the shifts
$b_{[1,2]}$ and the related charges $F_{[1,2]}$. This may be
achieved, say, by their hyperbolic re-parametrization with
$F(t)=\sqrt{2}\,e^t$ and $\cosh t = \sqrt{N}\,\cosh  \alpha(t)$ etc.
In terms of the parameters $t=t_{N,k}$ and/or $\alpha=\alpha_{N,k}$,
the spectrum of energies becomes deformed by the induced parametric
dependence of $b=b_{[1,2]}(t)=\pm\sqrt{2\cosh^2t-2N}
-e^{-t}/\sqrt{2}\,$. Purely formally, the eliminaton of $N$ might
extend the use of closed formulae up to~${\it L}=9$.

\section*{Higher partial waves}

\subsection{Numerical methods}

A shortcoming of the present SES construction lies in the growth of
its complexity at the large ${\it L}$. At ${\it L} \geq 5$ one
already cannot generate closed formulae for eigencharges, and the
purely numerical search for the roots $F$ is necessary at all the
very large ${\it L} \gg 1$. In an illustration using ${\it L}=3$, and
$b=5$ one gets the three eigencharges
\[
F = \{10.757,\ -10.400, \ -0.35755 \}
 \]
at the smallest possible $N=2$. They smoothly grow to the values
 \begin{equation}
 F = \{89.98,\ -89.975, \
-0.0049407 \} \label{largen}
 \end{equation}
evaluated at very large $N=1000$. This type of calculation is very
quick and its results exhibit a smooth $N-$dependence sampled in
Table~2.

\begin{table}[ht]
\caption{ $N-$dependence of eigencharges at $ {\it L}=4$ and $b=5$
}\label{table1}
\renewcommand\arraystretch{1.5}
\noindent\[
 \begin{array}{||c||cccc||} \hline
\hline
 N&\multicolumn{4}{c||}{F_{N,k}}\\
 \hline
 3&
  -15.\,\allowbreak
611&  -5.\,\allowbreak927\,9 &
 4.\,\allowbreak888\,7&16.\,\allowbreak651\\
 30&
-27.\,\allowbreak
149&-9.\,\allowbreak290\,9&8.\,\allowbreak929\,4&27.\,\allowbreak511\\
 300&
-74.\,\allowbreak 856&
 -24.\,\allowbreak984&24.\,\allowbreak936&74.\,\allowbreak904\\
3000&
 -232.\,\allowbreak82&-77.\,\allowbreak 610&
 77.\,\allowbreak605&232.\,\allowbreak83\\
 30 000&
 -734.\,\allowbreak99&
 -245.\,\allowbreak00&
 245.\,\allowbreak 00&734.\,\allowbreak99\\
 \hline
 \hline
\end{array}
\]
\end{table}

\subsection{Large $N$ expansions}

Any information concerning the eigenvalues $F=F(b)$ and/or $b=b(F)$
may shorten the necessary computations. At the intermediate ${\it
L}=5$ (which corresponds to the $d-$wave in three dimensions) we may
eliminate, for example, the product ${512\,F}N^\pm$ which is equal to
 \[ \left [ -768b-256Fb^{2}+768F+40F^{3}\pm 24\sqrt{\left(
1024b^{2}+192bF^{3}+512F^{2}+F^{6}\right) } \right ]
 \]
and use this formula as a constraint. Fortunately, closed formulae of
this type may also open the door to the perturbative methods.

Empirically, many numerically computed roots $F_{N,k}$ are very
smooth functions of $N$, especially in the asymptotic domain where
they may be well approximated by their available large$-N$ estimates.
For example, the exact result (\ref{largen}) already lies very close
to its leading-order analytic estimate
 \[
 F \approx
\{\sqrt{8N},\ -\sqrt{8N}, \ -b/N \} \approx \{89.44,\ -89.44, \
-0.005   \}. \label{lar}
 \]
At the intermediate values of the dimension $N$, precision may be
still insufficient but one can easily evaluate the large$-N$
corrections. For illustration, let us return to eq. (\ref{ouh}) and
re-write it, in the case of its smallest root $F = -\hat{b}\,
\hat{F}/N\,$, as the strictly equivalent formula
 \[
 \hat{F} = 1 + \frac{\beta}{N^3} \,\hat{F}^3,
 \ \ \ \ \ \ \beta=\frac{\hat{b}^3}{8b},
 \ \ \ \ \ \ \hat{b} = \frac{b}{1+(b^2-1)/2N}\ ,
 \ \ \ \ {\it L} = 3\ .
 \]
It is suitable for iterations which represent our root as the
following power series,
 \[
  \hat{F} = 1 + \frac{\beta}{N^3}
  + 3\,\frac{\beta^2}{N^6}
  + 12\,\frac{\beta^3}{N^9}
  + 28\,\frac{\beta^4}{N^{12}} + \ldots\ .
 \]
Also the other two  ${\it L} = 3$ roots may be represented by the
similar asymptotic series.

At the higher partial waves the same method works with a comparable
efficiency. At ${\it L} =4$, for example, the use of the variable $M
= \sqrt{2N+b^2-2}>0$ compactifies secular equation (\ref{recupe}),
 \[
 \left ( F^2-M^2 \right )\,
 \left ( F^2-9\,M^2 \right ) = 36-48\,b\,F\ .
 \]
In accord with Table 2, its asymptotically dominant ${\it O} (M^2)$
part determines the four distinct leading-order asymptotics of $F
\sim \varrho\,M$ where $\varrho = \pm 1, \pm 3$. Once we re-normalize
$F = \varrho\,M\,\sqrt{1+R}$, the four exact charges obey the
relation
 \[
 R =  {\frac{
 {48b \sqrt{1+R}}
 }
 {
  8|\varrho|-16 + \varrho^2R
 \,
 }}\ \frac{1}{
 \varrho\,M^3} +
 {\frac{
 {36}
 }
 {
  8|\varrho|-16 + \varrho^2R
 \,
 }}\ \frac{1}{
 \varrho^2 M^4}\ , \ \ \ \ {\it L} = 4\ .
 \]
Its iterations evaluate the correction term $R$ with an astounding
efficiency.

\specialsection{Quasi-variational solvability on model spaces}

\section*{``Sufficiently large" finite subspaces in Hilberet space}

\subsection{Left and right SES solutions}

A full-fledged applicability of the standard sets of QES wave
functions is weakened by their incompleteness. Moreover, all the
Hermitian QES Hamiltonians are usually interpreted as possessing just
a finite multiplet of elementary bound states (at a given, special
QES coupling) or Sturmians (cf. their example in Appendix B). For
this reason, even the study of their own small perturbations is not
easy at all \cite{perq}. Still, the sets of QES solutions themselves
become large enough when we take into consideration {\em both} their
fixed-coupling and fixed-energy subsets. In practice, such a
philosophy proved useful for approximation purposes \cite{Burrows}.

In this section, we are going to return to the latter idea in the
present broader, ${\it PT}$ symmetric SES context. For the sake of
definiteness let us  re-consider Hamiltonian of eq. (\ref{SE}) as if
it were defined at {\em any} charge $F$ standing at the Coulomb-like
force $W(r) = i\, r^{-1}$. Then we may denote $H = H(F)= H(0) + F\,W$
and, in the spirit of the Dirac's bra and ket notation, abbreviate
$|\psi_{M,k}^{(-)}\rangle= |M,k\rangle$. This enables us to re-write
the special SES version of eq. (\ref{SE}) in shorthand,
 \begin{equation}
  H(0)
 \ |M,k\rangle
  +W\ |M,k\rangle\, F_{M,k}
 =  |M,k\rangle\,E_M
 \label{one}
 \end{equation}
 \[
 \ \ \ \ \ \ \ \ \ \ \  \ \ \ \
 \ \ \ \ \ \ \ \ \ \ \  \ \ \ \
 \ \ \ \ \ \ \ \ \ \ \  \ \ \ \
 k =  1, \ldots, {\it L}, \ \ \ M = 0, 1, \ldots \ .
 \]
Our ${\it PT}$ symmetric Hamiltonian is non-Hermitean so that its
left and right eigenstates will differ in general. One has to
complement eq. (\ref{one}) by its counterpart
 \begin{equation}
 \langle \langle N,j | \
 H(0) + F_{N,j}\, \langle \langle N,j | W
 = E_N\ \langle \langle N,j |,
 \label{two}
 \end{equation}
 \[
 \ \ \ \ \ \ \ \ \ \ \  \ \ \ \
 \ \ \ \ \ \ \ \ \ \ \  \ \ \ \
 \ \ \ \ \ \ \ \ \ \ \  \ \ \ \
 j =  1, \ldots, {\it L}, \ \ \ N = 0, 1, \ldots \
 \]
where the same operators act to the left. After a return to the
differential form of this problem (\ref{SE}), a transition between
Hermitian-conjugate equations (\ref{one}) and (\ref{two}) may be
re-interpreted as a certain reflection ${\it R}$ in the space of the
parameters,
 \begin{equation}
 {\it R} \ : \ \left\{
 F\  \longleftrightarrow\  -F,\ \ \
 b\ \longleftrightarrow\ \, -b, \ \ \
 \varepsilon\  \longleftrightarrow\ \ -\varepsilon\,.
 \right \}.
 \label{reflexe}
 \end{equation}
In an illustration using $L=2$, our SES states of section 1 may be
represented by a discrete set of points in the energy-charge plane.
The values of their coordinates $(E_N,F_{N,k} )$ are specified by the
respective {\em closed formulae} (\ref{Hautots}) and (\ref{refex}).
These points may be perceived as located on the two (viz. left and
right) branches of a single hyperbolic curve. These two {infinitely
large} multiplets are mutually related by the transformation ${\it
R}$ of eq. (\ref{reflexe})so that ${\it R}|N,1\rangle =
|N,2\rangle\rangle$ and {\it vice versa}.

\subsection{Generalized Sturmian SES states as a basis}

There is no {\it a priori} reason for a bi-orthogonality between our
multi-indexed bras $ \langle \langle N,k|\ \equiv\ \langle \langle
A|\ $ and ket vectors $\ |N',k'\rangle\ \equiv\   |b\rangle \ $ so
that their overlaps
 \[
 Q_{A,b}= \langle \langle A | b \rangle
 \]
form a non-diagonal and asymmetric matrix in general. We have to
assume that after a finite-dimensional truncation, this overlap
matrix is invertible. Only in such a case we may define the inverse
$R = Q^{-1}$ and introduce an identity projector in a ``sufficiently
large" subspace of Hilbert space,
 \begin{equation}
 I = \sum_{
 a \in {\it J}_{ket},
 B \in {\it J}_{bra}
 }
 |a\rangle\,R_{a,B}\,\langle \langle B|\ .
 \label{projek}
 \end{equation}
In a constructive mood, let us return to equations (\ref{one}) and
(\ref{two}) and imagine that they share all their eigen-energies and
eigen-charges. This enables us to write down the following two
alternative matrix equations
 \begin{equation}
 \langle \langle N,j | \
 H(0)
 \ |M,k\rangle = \langle \langle N,j |M,k\rangle\  E_{M}
 -
 \langle \langle N,j | \
  W
 \ |M,k\rangle\ F_{M,k}
 \label{oned}
 \end{equation}
 \begin{equation}
 \langle \langle N,j | \
  H(0)
 \ |M,k\rangle = E_{N}\  \langle \langle N,j |M,k\rangle
 - F_{N,j}\,
 \langle \langle N,j | \
  W
 \ |M,k\rangle
 \label{twod}
 \end{equation}
with $ (N,j) \in {\it J}_{bra}$ and $ (M,k) \in {\it J}_{ket}$. Their
subtraction gives the constraint
 \begin{equation}
 \left (
 F_{M,k}-F_{N,j}
 \right )\,
 \langle \langle N,j | \
  W\ |M,k\rangle
  =
 \left (
 E_{M}-E_{N}
 \right )\,
  Q_{({N,j}),({M,k})}\,
  .
 \label{biogg}
 \end{equation}
Due to the way of its derivation, this relation may be understood as
an immediate generalization of bi-orthogonality. In particular, we
see that within the subspace of a single Sturmian multiplet (i.e.,
for $M = N$), the left-hand side expression must be a diagonal matrix
with respect to its second indices. For our present purposes we shall
abbreviate $\langle \langle N,j |  W |N,j\rangle \equiv w_{N,j}$.

\section*{Non-QES bound-state problems}

\subsection{Matrix equations}

The knowledge of a basis can facilitate the study of many perturbed
Hamiltonians via textbook perturbation recipes \cite{Landau}. The use
of the present SES bases might lead to a new progress in this area.
Let us recall the respective right and left forms of any
Schr\"{o}dinger non-QES bound state problem with the above-mentioned
structure,
 \[
 \left [ H(0) + F\,W
 \right ]\ |\Psi \rangle = E(F) \ |\Psi \rangle\
  \label{obone},
 \]
 \[
 \langle \langle \Psi |\
 \left [ H(0) + F\,W
 \right ]= E(F) \
 \langle \langle \Psi |\
  \label{obtwo}.
 \]
Assuming that $ F\neq F^{(QES)}$ and using eq. (\ref{projek}) we may
insert
 \[
 |\Psi\rangle= \sum_{
 a \in {\it J}_{ket},
 B \in {\it J}_{bra}
 }
 |a\rangle\,R_{a,B}\,\langle \langle B|\Psi \rangle
 = \sum_{
 a \in {\it J}_{ket}
 }
 |a\rangle\,h_{a}\,
 \ ,
 \]
 \[
 \langle \langle \Psi |
  = \sum_{
 a \in {\it J}_{ket},
 B \in {\it J}_{bra}
 } \langle \langle \Psi
 |a\rangle\,R_{a,B}\,\langle \langle B|
 = \sum_{
 B \in {\it J}_{bra}
 } {g}_{B}\,\langle \langle B|\,
 \
 \]
and arrive at the double eigen-problem
 \[
 \sum_{b \in {\it J}_{ket},
 C \in {\it J}_{bra}
 }
 \,\langle \langle A | \,H(F)\,|b\rangle\,
 R_{b,C}\,\langle\langle C|\Psi\rangle = E\,
 \langle\langle A|\Psi\rangle, \ \ \ \ \ \ \ \ \ A \in {\it
 J}_{bra}
 \]
(plus its -- omitted -- conjugated companion), i.e., at the two
conjugate linear algebraic systems of equations written in terms of
the same matrix,
 \begin{equation}
 {\bf Z}(E,F) \,\vec{h} = 0, \ \ \ \ \ \
 \vec{{g}}^{\dagger}\,{\bf Z}(E,F)
  = 0
  \label{dvoje}
 \end{equation}
 \[
 {\bf Z}_{A,b}(E,F)
 = \langle \langle A |\ H(0)\ | b \rangle
 -E\, \langle \langle A | b \rangle
 +F\, \langle \langle A |\ W\ | b \rangle\ .
 \]
Now, one could introduce a small parameter $\lambda= F- F^{(QES)}$
and try to construct, say, $E= E(\lambda)$ in the form of a power
series in $\lambda$.

\subsection{Matrix elements}

Long before any numerical determination of bound states, we must
evaluate {\em all} the necessary matrix elements as an input. In
practice, the latter step is usually the most time-consuming part of
the algorithm. One has to optimize it in the present QES setting,
therefore.

In the first step we recall eq. (\ref{twod}) and eliminate all the
matrix elements of $H(0)$. This means that in eq. (\ref{dvoje}) we
reduce the costly input to the mere evaluation of the matrix elements
of the weakly singular Coulomb potential $W(r)=i/r$,
 \[
 {\bf Z}_{A,b}(E,F)
 =
  \left ( F-F_A \right ) \, \langle \langle A |\ W\ | b \rangle
 -\left ( E-E_A \right )\, \langle \langle A | b \rangle
  \ .
 \]
In the second step we keep $M \neq N$ (i.e., we stay out of the
Sturmian subspaces or diagonal blocks in the matrix ${\bf Z}$) and
postulate the absence of a random degeneracy of charges. This means
$F_{M,k} \neq F_{N,j}$ so that we are permitted to re-arrange the
bi-orthogonality-like relation (\ref{biogg}) into a further reduction
of the necessary input information,
 \[
 \langle \langle N,j | \
  W\ |M,k\rangle
  =
 \frac{
 E_{M}-E_{N}
 }{
 F_{M,k}-F_{N,j}
 }
  Q_{({N,j}),({M,k})}\,, \ \ \ \ \ \ \ \ \ \ M \neq N\ .
 \label{bioggpp}
 \]
In this way we arrive at the final form of our linear Schr\"{o}dinger
non-QES algebraic problem for the right eigenvectors,
 \[
 w_{N,j}\,h_{N,j}+
 \sum_{K(\neq N),p}\,
  \frac{E_N-E_K}{F_{N,j}-F_{K,p}}\
 Q_{(N,j),(K,p)}\,h_{K,p}   =
 \]
 \[
  \ \ \ \ \ \ \ =
 \frac{
 E-E_N
 }{
 F-F_{N,j}
 }\,
 \  \sum_{M,k}\,
 Q_{(N,j),(M,k)}\,h_{M,k}\ ,
 \]
 \[
 \ \ \ \ \ \ \ \  \  \ \ \ \ \ \
 \ \ \ \ \ \ \ \
 \ \ \ \ \ \ \ \  \
 \ \ \ \ \ \ \ \  \
 j = 1, 2, \ldots, {\it L}, \ \ \ \
 N = 0, 1, \ldots\ .
 \]
For left eigenvectors, the system of equations is very similar though
not equivalent,
 \[
 w_{N,j}\,{g}_{N,j}+
 \sum_{K(\neq N),p}\,
  \frac{E_N-E_K}{F_{N,j}-F_{K,p}}\
 Q_{(K,p),(N,j)}\,{g}_{K,p}   =
 \]
 \[
  \ \ \ \ \ \ \ =
 \frac{
 E-E_N
 }{
 F-F_{N,j}
 }\,
 \  \sum_{M,k}\,
 Q_{(M,k),(N,j)}\,{g}_{M,k}\ ,
 \]
 \[
 \ \ \ \ \ \ \ \  \  \ \ \ \ \ \
 \ \ \ \ \ \ \ \
 \ \ \ \ \ \ \ \  \
 \ \ \ \ \ \ \ \  \
 j = 1, 2, \ldots, {\it L}, \ \ \ \
 N = 0, 1, \ldots\ .
 \]
To solve any of these systems, say, by a perturbation technique, we
just need to know the overlaps $Q$ and the vector of Coulombic
elements $w_{N,j}$.

\specialsection{Concluding questions and remarks}

\subsection{Does the quantum quasi-exact solvability
have a classical analogue?}

In some studies\footnote{as reviewed, e.g., by Pavel Winternitz in
this volume \cite{PW}} a new relationship has been traced between
classical and quantum mechanics. Its core may be seen in a
correlation between the concepts of {\em integrability} and {\em
solvability}. Needless to repeat, the former feature plays a key role
in classical systems while its versions known as superintegrability
and maximal superintegrability acquire more relevance in quantized
world.

For many quantum Hamiltonians the latter qualities also seem related
to a purely algebraic property of quasi-exact solvability
\cite{Rodriguez}. In such a context, it is extremely exciting to ask
questions about the robustness of the latter correlation {\em and}
about the natural ways of the definition of the solvability itself.
In a purely pragmatic manner, ${\it PT}$ symmetrization could offer
here another bridge towards the new solvable models. It is evident
that after one weakens the traditional requirement of Hermiticity,
many new QES models may be found, indeed \cite{sextic,decadic}.

In the present paper we have seen that for the $D$ dimensional
central Coulomb plus shifted harmonic oscillator its quasi-exact
solvability acquires a fairly unusual modified form. A significant
difference appears between the quasi-even and quasi-odd states, where
only the properties of the latter set remain standard. The
facilitated construction and unexpectedly non-numerical character of
the former quasi-even family make it similar to the current complete
oscillator basis on half-axis. This locates our new ``semi-exact"
solutions in a gap between their older exact and quasi-exact
neighbors.

\subsection{Do we need more semi-exactly solvable models?}

Our presentation of details started from the reduction of the
four-term recurrences of our ``paper I" \cite{I} to their improved
three-term form. This was rendered possible by our new ansatz which
also proved able to reproduce all the available older results. We
underlined that one of the most characteristic shortcomings of the
standard QES equations lies in the necessity of solving the adjacent
linear algebraic $N-$dimensional eigenvalue problem of growing size
$N$ which, in our particular example, selects the admissible charges.

In this context it is important that we succeeded in revealing the
existence of a new, ``facilitated" or ``semi-exact" elementary
solvability emerging when the quasi parity was assumed even.  {\em
Infinitely many} of these states acquire the exact polynomial form in
each (= $m-$th) partial wave and at any element $E=E(N)$ of an
equidistant set of the energies.  In fact, our states form the
Sturmian ${\it L}-$plets (numbered by $N = 0, 1, \ldots$) at charges
$F_{N,k}$ numbered by $k = 1, 2, \ldots, {\it L}\ $ where ${\it L} =
D-2+2m$ is $N-$independent.

The well known even-parity basis on $L_2(0,\infty)$ (made of Hermite
polynomials) is re-obtained in the simplest ${\it L} =1$ special case
where the resulting (single) QES charge is zero, $F=0$.  In the first
nontrivial ${\it L} =2$ case we have got the two alternative eligible
infinite series of states at charges $F =\pm\sqrt{ b^{2} +2N}$ (where
$b$ is the shift). We have shown that and how the similar sets might
serve as a source of the new matrix reformulations and approximation
techniques within perturbation theory or variational considerations.

Our main attention has been paid to the generalized QES ansatz
admitting solutions with both quasi-parities. Our main result, viz.,
a new version of the SES recurrent construction is ``almost exact",
first of all, due to its $N-$independence. In the first few lowest
partial waves our infinite families of semi-exact states proved very
transparent and close, by their applicability, to the basis of
harmonic oscillators. We revealed that their construction is amenable
to an efficient approximative treatment even at the higher partial
waves via the Taylor-series expansions similar to the common
large$-N$ perturbation theory. This might not only initiate the study
of the old models in the non-QES domains of couplings but also offer
a new motivation for an intensified search for the new SES models.

\subsection{Will our new SES bases find efficient practical applications?}

The matrices of overlaps $Q_{A,b}$ are, presumably, non-diagonal. In
the other words, our SES basis states $ |N,k\rangle$ and $|N,k\rangle
\rangle$ are {\em not} mutually bi-orthogonal. Any deeper insight in
their overlaps would be appreciated in applications, therefore. In
the future, this could facilitate, say, a search for a non-Hermitian
analogue of the QES + extrapolation trick as suggested by Burrows et
al \cite{Burrows} for the Hermitian sextic anharmonicities.

Our formalism is not yet fully prepared for initiation of any
practical numerical or perturbative calculations in the non-QES
regime of course. At the same time, we already succeeded in a drastic
reduction of the number of the necessary input matrix elements. This
supports our belief that our present constructions represent a
valuable key step towards deeper understanding of relations between
the degree of integrability in classical mechanics and an extent of
exact solvability in quantum mechanics. At this moment we have to
re-emphasize the definite progress in our understanding of properties
of the ``forgotten" quasi-even solutions, their quasi-exact
solvability of which may be characterized as ``significantly
facilitated". They are distributed in the coupling-energy plane along
curves which may be interpreted as a common generalization of the
usual bound-state straight lines\footnote{characterized by the fixed
couplings and variable energies and exemplified by the spectrum of
the exactly solvable harmonic oscillator or by the Singh's sextic QES
oscillators \cite{Singh}} and the straight lines of the so called
Sturmian states\footnote{characterized by the variable couplings and
fixed energies and exemplified by the exactly solvable zero-order
models in ref. \cite{perS} or by the Flessas' non-polynomial QES
oscillators \cite{Flessas}}. In this sense, their families may be
understood as certain generalized Sturmian families defined on some
curves in the plane of charge and energy. These sets contain
infinitely many elements and definitely resemble and generalize the
current types of bases.

\subsection{Could the complexification
serve as a regularization recipe for SW models?}

In connection with the strongly singular behaviour of many
classically superintegrable models at $r=0$, a new role of the
present weakening of their Hermiticity may be also sought in its
parallel regularization effect. In this role, the ${\it PT}$
symmetric regularization proved very useful and really vital in the
various ${\it PT}$ symmetrized versions of the so called
supersymmetric (SUSY) quantum mechanics \cite{Cannata,Bagchi}. For
example, in accord with refs. \cite{ptsusy} the SW Hamiltonians $H=
p^2+r^2+G/r^2$ may form the {\em formal} SUSY partners $H_{(L,R)}$.
Once their singularity at $r=0$ is circumvented via the complex shift
of the coordinate axis, their spectra remain real and discrete and
exhibit still the usual SUSY-isospectrality pattern in the so called
unbroken ${\it PT}$ symmetry domain with $G>-1/4$.

In  \cite{Paris} we have shown that beyond the latter domain, the
${\it PT}$ symmetry itself is completely broken but the spectra stay
partially real on a SUSY-generated set of complex $G$. In addition,
an alternative SUSY scheme with equal spikes, $G_{(L)} = G_{(R)}$ may
be introduced giving an innovated super-Hamiltonian factorized in
terms of certain creation and annihilation differential operators of
the second order which conserve the quasi-parity and are mutually not
adjoint. Together with the original Hamiltonian $H$ the latter two
operators commute in accord with the Lie algebra~$sl(2,I\!\!\! R)$.

All these preliminary results and hints are encouraging and suggest
that in the context of the Calogero and/or Smorodinsky-Winternitz
superintegrable models with strong barriers the ${\it PT}$
symmetry-induced tunneling effects did not say their last word yet.

\bibliographystyle{amsalpha}

\specialsection{Appendices}

\section*{A. Charged harmonic oscillator and the concept of
quasi-parity}

As we already mentioned, the shifted and charged harmonic oscillator
of ref. \cite{Hautot} is a characteristic illustrative example of QES
model in quantum mechanics. Its Hermitian Schr\"{o}dinger equation
\begin{equation}
\left[-\,\frac{d^2}{dr^2}+ \frac{\ell(\ell+1)}{r^2}+ \frac{f}{r} +
2\,g\,r + r^2\right]\,
 \varphi(r) = E\, \varphi(r), \ \ \ \ \ r \in (0, \infty)
 \label{SEH}
 \end{equation}
has been replaced by its ${\it PT}-$symmetrized version (\ref{SE}) in
ref. \cite{I}. After the severe restriction of the potential to
$s-$waves in three dimensions (i.e., to $D=3$ and $\ell = 0$,
motivated by technical reasons) we succeeded in confirming its
quasi-exact solvability in non-Hermitian regime. In the present
continuation of paper \cite{I} this result is extended to all the
partial waves $\ell = (D-3)/2, (D-3)/2+1, \ldots$ and dimensions $D =
3, 4, \ldots$.

The first hint indicating the open possibilities in connection with
eq. (\ref{SEH}) appeared in ref. \cite{ptho} where the concept of
quasi-parity has been introduced. One was able to conclude there that
all the solutions of the differential Schr\"{o}dinger equations of
the radial type prove separated into quasi-even or quasi-odd states
$\psi_n^{ (\pm)} (x)$ in a way which complements the quasi-odd
solutions of ref. \cite{I} by their present ``forgotten" partners
with quasi-even symmetry. The key notion of the quasi-parity itself
may be defined either via asymptotics or, more easily, via the
specific behaviour of the wave functions near the singularity at
$r(x) \sim 0$,
 \[
 \psi_n^{(+)}(x) \sim (x - i\,\varepsilon)^{-\ell}, \ \ \ \ \
 \psi_n^{(-)}(x) \sim (x - i\,\varepsilon)^{\ell+1}, \ \ \ \ \
  x \sim i\,\varepsilon.
  \]
The name suggests that the quasi-parity coincides with the ordinary
parity  after limiting transition $\varepsilon \to 0$ to a Hermitian
Hamiltonian. In this limit, the quasi-even solutions themselves need
not lose their normalizability (and may remain physical) whenever the
strength of their singular repulsion is sufficiently small (cf. an
explicit illustration in ref. \cite{ptsusy}).

\section*{B. QES states with odd quasi-parity}

Formula (\ref{Hautots}) for the energies is independent of the charge
and was already known to Hautot \cite{Hautot}. Thus, the energy is a
constant for all the solutions of eq. (\ref{recu}). The multiplets of
bound states of such a type\footnote{defined at $N$ different charges
$F_{N,j}$ and called Sturmians} find applications, say, in
perturbation theory \cite{perS}. Up to a free normalization $p_N\neq
0$, their present closed definition
 \begin{equation}
 p_{j-1}^{(-)}
= {{p_N^{(-)}}\over {(-A_j)(-A_{j+1}) \ldots (-A_{N})}}
 \,\det
 \left( \begin{array}{ccccc} B_{j} & C_j&  & &  \\
A_{j+1}&B_{j+1} & C_{j+1}&    & \\
 &\ddots&\ddots&\ddots&\\
&&A_{N-1}&B_{N-1}&C_{N-1}\\
 &&&A_{N}&B_{N}
\end{array} \right)
 \label{exrecu}
  \end{equation}
is unique at any energy $E$, shift $b$ and charge $F$.

Our new ansatz (\ref{ansatz}) is sufficiently flexible and reproduces
all the older quasi-odd QES solutions of refs. \cite{Hautot} and
\cite{I}. Indeed, we may mark them by the superscript $^{(-)}$ and
characterize their set by the following boundary condition,
 \begin{equation}
p_0^{(-)}=p_1^{(-)}=\ldots = p_{{\it L}-1}^{(-)}=0\ , \ \ \ \ \
 p_{{\it L}}^{(-)}\neq 0\
\label{condi}
 \end{equation}
which reduces the range of the indices in formula (\ref{exrecu}), $j
= N, N-1, \ldots, {\it L}\ $. We have to guarantee that the QES
recurrent recipe terminates, i.e., that the value of $p_{{\it
L}-1}^{(-)}$ vanishes. This condition has the secular form
 \begin{equation}
  {\it S}^{(-)} =\det
 \left( \begin{array}{ccccc} B_{{\it L}}{}
  & C_{{\it L}}{}
 &  & &  \\
 A_{{\it L}+1}{}&B_{{\it L}+1}{} & C_{{\it L}+1}{}&
 &
 \\
 &\ddots&\ddots&\ddots&\\
 &&A_{N-1}{}
 &B_{N-1}{}&C_{N-1}{}\\
 &&&A_{N}{}&B_{N}{}
 \end{array}
  \right)\ = 0
 \label{qesrec}
 \end{equation}
and its occurrence is characteristic for all QES systems
\cite{Ushveridze}. At every admissible energy it specifies a
multiplet of the admissible charges $F=F^{(-)}_{N,j}$ numbered by the
second subscript $j = 1, 2, \ldots, N-{\it L} +1$. At each $N$ and
${\it L}$ our secular polynomial is of degree $N-{\it L}+1$. Of
course, the practical determination of the quasi-odd QES eigencharges
is a purely numerical procedure for $N>{\it L}+3$. Fortunately, as we
have seen in section~1, this shortcoming may be suppressed in the
quasi-even case.

\end{document}